\documentclass[aps, twocolumn, prl, 10pt]{revtex4-1}

\usepackage{graphicx}
\usepackage{dcolumn}
\usepackage{bm}
\usepackage{sidecap}
\usepackage{amssymb}
\usepackage{amsmath}
\usepackage{color,xcolor}

\definecolor{MLgreen}{rgb}{0,0.7,0}

\newcommand{\blu}{\color{black}}

\DeclareGraphicsExtensions {.pdf,.png,.jpg,.eps}

\begin{document}

\title {Interference of short optical pulses from independent gain-switched laser diodes\\
{\blu for quantum secure communications}}

\author{Z.~L.~Yuan}
\email{zhiliang.yuan@crl.toshiba.co.uk}
\author{M.~Lucamarini}
\author {J.~F.~Dynes}
\author {B.~ Fr\"ohlich}
\author {M.~B.~Ward}
\author {A.~J.~Shields}
\affiliation{Toshiba Research Europe Limited, Cambridge Research Laboratory, 208 Cambridge Science Park, Milton Road, Cambridge, CB4~0GZ, United Kingdom}
\date{\today}

\begin{abstract}
Since the introduction of the decoy-state technique, phase-randomised weak coherent light pulses have been the key to increase the practicality of quantum-based communications.
{\blu Their ultra-fast generation was accomplished via compact gain-switched (GS) lasers, leading to high key rates in quantum key distribution (QKD).}
{\blu Recently, the question arose of whether the same laser could be employed to achieve high-speed} measurement-device-independent-QKD, a scheme that promises long-haul quantum communications immune to all detector attacks. {\blu For that, a challenging high-visibility interference between independent picosecond optical pulses is required.}
{\blu Here, we answer the above question in the {\blu affirmative by demonstrating} high-visibility interference from two independent GS lasers triggered at 1GHz.} The result is obtained through a careful characterization of the laser frequency chirp and time jitter. {\blu By relating these quantities to the interference visibility, we obtain a parameter-free verification of the experimental data and a numerical simulation of the achievable key rate{\blu s. T}hese findings are beneficial to other applications making use of GS lasers}, including random number generation and standard QKD.
\end{abstract}

\maketitle

Interference lies at the heart of quantum information technologies. Novel protocols {\blu and schemes}, such as quantum cryptography~\cite{gisin02}, quantum teleportation~\cite{bouwmeester97}, quantum {\blu repeaters}~\cite{briegel98},
or linear optics quantum computing~\cite{knill01} rely upon high visibility interference of {\blu light pulses}. To achieve high visibility, {\blu the interfering pulses} need to be indistinguishable in all possible degrees of freedom~\cite{hong87,rarity05,kaltenbaek06,beugnon06,bennett09}.

\begin{figure}[tbp]
\centering
\includegraphics[width=7.5cm, height=5cm]{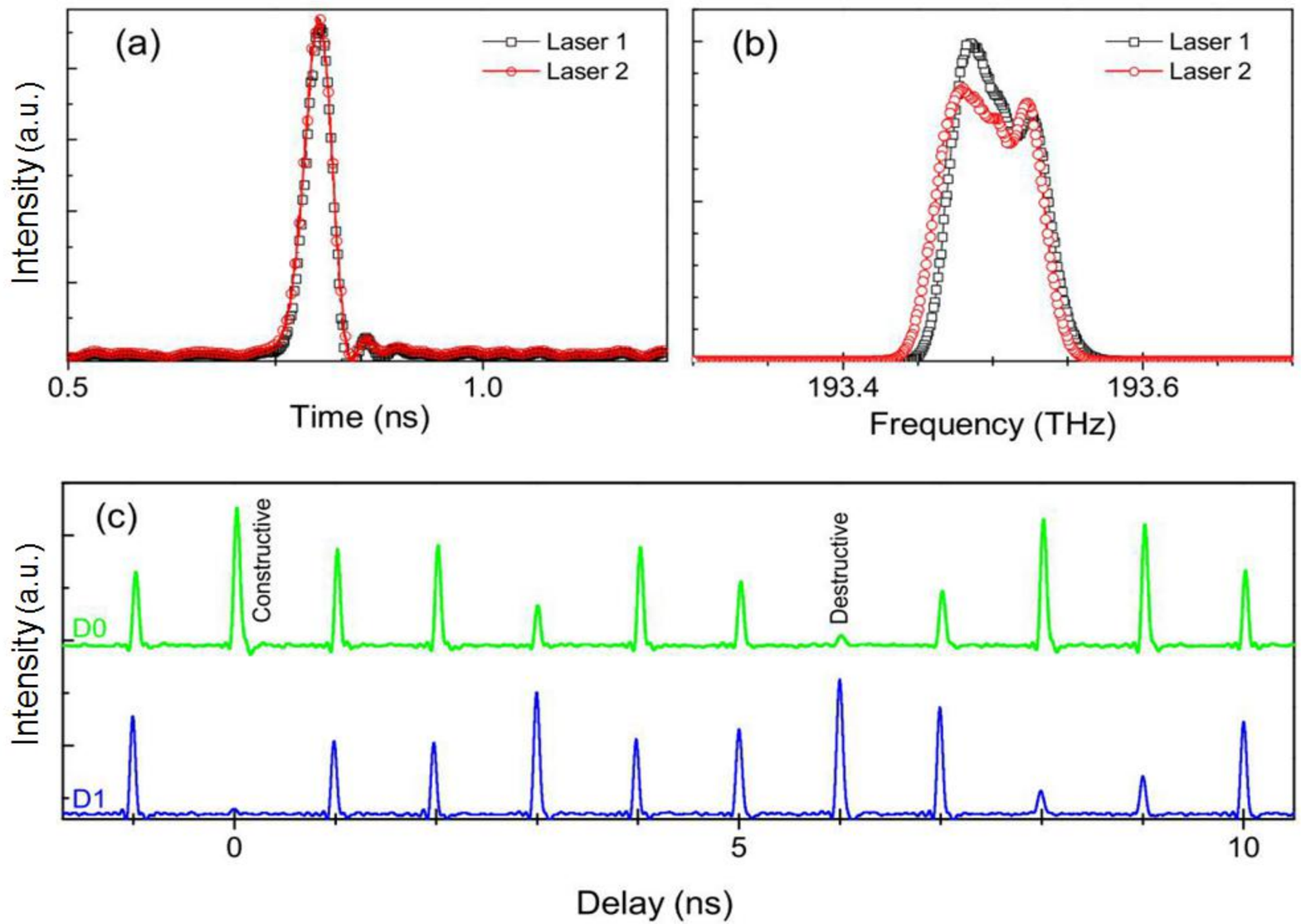}
\caption{(a) {\blu Temporal} and (b) {\blu spectral} profiles of the two GS lasers used in this work's experimental apparatus. (c) Interference output traces recorded by an oscilloscope and two fast photodiodes (D0 and D1).}
\label{fig:fig1}
\end{figure}

{\blu Weak coherent states of light} have been {\blu long} used to approximate single photon sources in quantum key distribution (QKD). {\blu This approximation guarantees high key rates if} the decoy{\blu -state} technique~\cite{lo05,wang05} {\blu is adopted. However, in order to apply it, the electromagnetic phase of coherent states is required to be random}.
Thankfully, semiconductor gain-switched (GS) laser diodes naturally generate optical pulses with random phases~\cite{yuan14,kobayashi14}. With a sufficient off period between subsequent events, each laser pulse is triggered by quantum-mechanical spontaneous emission and thus has random electromagnetic phase~\cite{yuan14}. At the same time, GS short pulses (around 30~ps, see Fig.~\ref{fig:fig1}(a)) are perfectly suited  to high bit rate~\cite{lutkenhaus09,sasaki11} and noise-tolerant QKD~\cite{patel12prx}. {\blu This is remarkable given that time-jitter in GS lasers is about 10~ps, i.e., comparable to the pulse width.} {\blu Furthermore, other potential sources of impairment like the pulses spectral distinguishability or a time-varying polarization, play only a minor role in standard QKD, where each generated pulse only interferes with itself to deliver a bit of the final key.}

{\blu The situation is dramatically {\blu changed} in measurement-device-independent-QKD (MDI-QKD)}, a recent quantum protocol promising immunity against all detector attacks~\cite{lo12,silva13,rubenok13,liu13,tang13mdi}. Similarly to conventional QKD, {\blu decoy states and phase randomization are} {\blu also} required in MDI-QKD. However, {\blu in this case, the successful distillation of the final key requires high visibility two pulse interference}~\cite{lo12}. {\blu This poses stringent requirements on the system, as the interfering pulses have to be indistinguishable and perfectly overlapped to guarantee high visibility.} {\blu Time jitter and frequency profile of the pulses} {\blu play a {\blu very} important role and it is {\blu unclear} whether GS lasers represent a viable solution}.

Until now it has not been possible to use GS pulses {\blu shorter than 2~ns and trigger rates higher than 1~MHz in an MDI-QKD experiment~\cite{liu13}. This is still orders of magnitude away from {\blu high bit-rate QKD\cite{sasaki11,dixon08}, working at 1~GHz} with pulse widths of tens of ps.}
{\blu The tolerance to time and frequency fluctuations could be improved by using the} steady-state emission of GS laser diodes~\cite{yuan14}. However, this would limit the prospect for high bit rate applications.
{\blu Other} MDI-QKD demonstrations~\cite{silva13,rubenok13,tang13mdi} {\blu have improved the spectral stability of the pulses} approximating the required source with continuous-wave (CW) lasers pulse-carved by an intensity modulator. Light pulses generated this way exhibit {\blu a negligible time-jitter, but also a} constant, or slowly variable, phase, therefore violating the phase randomness requirement. Phase randomisation via {\blu separate} modulation is possible~\cite{zhao07,tang13mdi}, but {\blu at the expense of} additional
complexity of the {\blu setup~\cite{error3}.}

Here, we investigate the relation between GS laser diodes and interference visibility {\blu and implement a solution to mitigate the detrimental effect {\blu of pulse distinguishability}. By introducing a novel theoretical model, we identify {\blu frequency chirp of GS pulses} as the main cause of poor interference visibility}.
{\blu Frequency chirp is common in fast-driven semiconductor laser diodes~\cite{agrawal02}.} {\blu The rapid change in carrier density in the active region} dynamically alters the refractive index thus chirping the laser frequency~\cite{pataca97} and {\blu making the pulses far from transform limited~\cite{yariv06} (see, e.g., Fig.~\ref{fig:fig1}(b)).}
The concomitant time jitter then prevents two chirped pulses from preserving a constant phase relation, which is {\blu a} prerequisite for high-visibility interference.
We verify this analysis by experimentally interfering short optical pulses emitted by two independent semiconductor GS laser diodes {\blu driven at 1~GHz} and {\blu comparing the results with the theoretical prediction.}
{\blu When frequency chirp is taken into account, the theory provides a parameter-free fit of the experimental data, thus confirming the soundness of our analysis.}
{\blu This fact was then exploited to calibrate our system and achieve two-pulse interference visibility as high as 0.46, close to the theoretical limit 0.50 achievable with weak coherent states~\cite{rarity05}. Combined with the intrinsic phase randomness of the pulses and the high trigger rate of the laser, this result demonstrates the usefulness of GS laser diodes in achieving {\blu high speed} decoy-state MDI-QKD.}
Furthermore, {\blu it} has implications for other types of {\blu high-speed} quantum information applications{\blu , as discussed later on}.

\begin{figure}[tbp]
\centering
\includegraphics[width=.88\columnwidth, height=3.75cm]{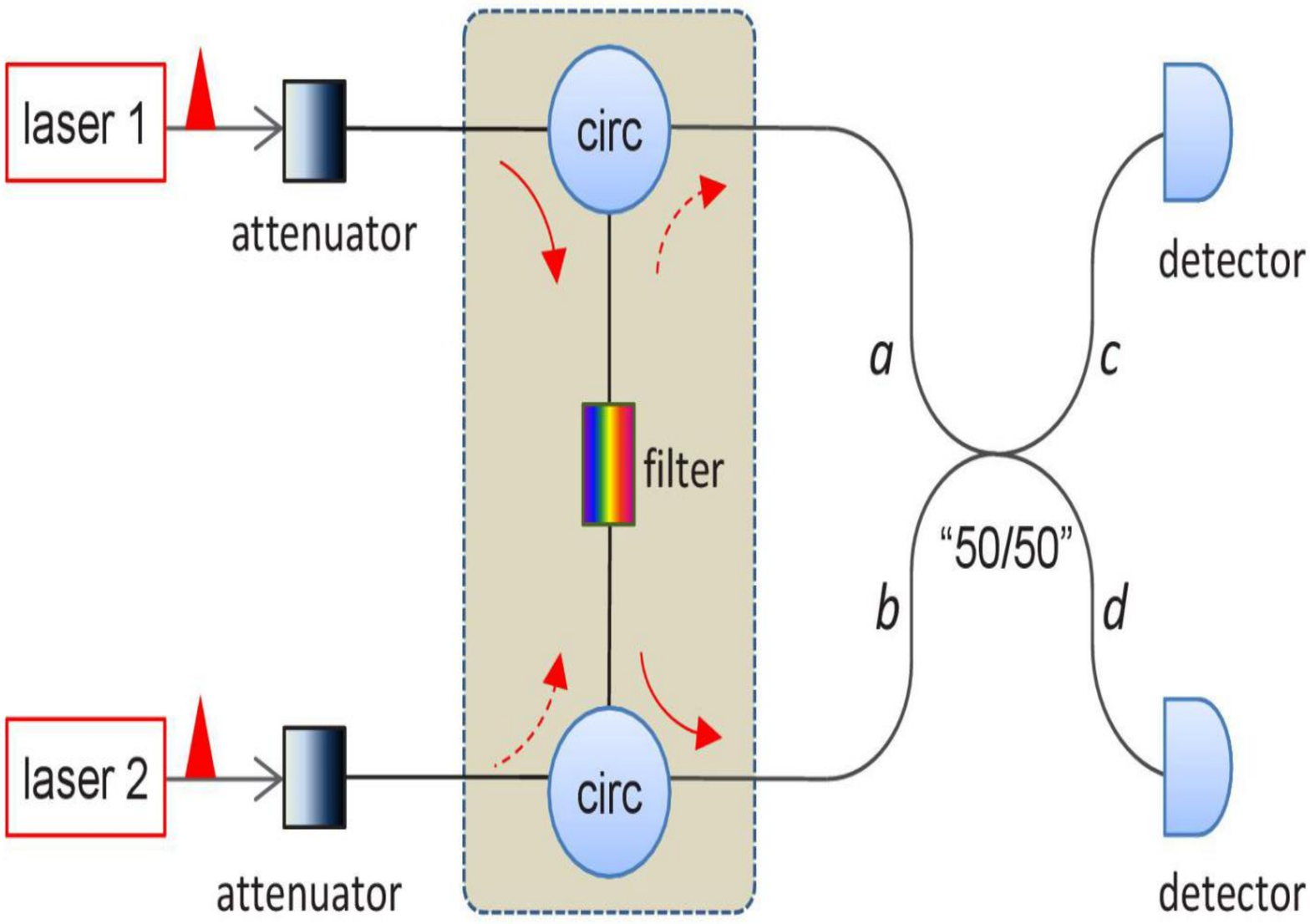}
\caption{{\blu Schematics of the} experimental setup. The shade illustrate the spectral filtering using a single filter with the help of two optical circulators. {\blu The same diagram is used for modelling the effect of time jitter and frequency chirp on the two-pulse interference visibility.}}
\label{fig:fig2}
\newpage
\end{figure}

We start our analysis by {\blu describing the schematics of} {\blu the} experimental setup, depicted in Fig.~\ref{fig:fig2}, {\blu based on which we develop our theoretical model.}
{\blu The setup} consists of a Hong-Ou-Mandel interferometer~\cite{hong87}, with two attenuated GS distributed feedback laser diodes {\blu injecting light into} a beam splitter {\blu through a pair of} optical circulators and a tunable filter. After the beam splitter, light is detected by two single photon detectors, {\blu thus emulating a real MDI-QKD setup}. Ideally, a tunable bandpass filter {\blu should appear in each arm of the setup} to limit the bandwidth and hence the frequency chirp of each laser. For {\blu experimental} convenience, we use a single tunable bandpass filter {\blu and two}  optical circulators to filter emissions of both lasers simultaneously{\blu ~\cite{error4}}. Each laser is attenuated by more than 70~dB {\blu up} to the single photon level before interference.  Including the built-in isolation (30 dB) in each laser diode and the circulator extinction ratio of 50~dB, the total isolation between the light sources is greater than 150~dB.  Considering each laser emitting an optical power of $\sim$200~$\mu$W at 1~GHz, this level of isolation ensures that the optical cross talk between the lasers is less than $10^{-8}$ photons/pulse. We therefore conclude the laser diodes are optically independent.

Gain-switching in the laser diodes is achieved electrically by a superposition of a DC bias and a voltage square wave clocked at 1~GHz. Temporal alignment of the pulses, shown in Fig.~\ref{fig:fig1}(a), is achieved by tuning the delay of Laser 2, which is electronically adjustable in steps of 1~ps. Laser~1 is kept at room temperature, while laser~2 is cooled to $-9^\circ$C to tune its central frequency (193.47~THz) to approximately match Laser 1, as shown in Fig.~\ref{fig:fig1}(b).   The second order correlation functions at 0-delay are measured for lasers 1 and 2 and amount to $99.3\pm1.0\%$ and $99.6 \pm 1.3\%$, respectively, suggesting the Poisson statistics as expected for coherent state emission \cite{dixon09}.
Optical pulses from both lasers exhibit temporal and spectral full widths at half maximum (FWHM's) of 30~ps and 70~GHz, respectively.  Excluding the influence from the time jitter, which was measured to be 9.3~ps (FWHM), the laser pulses are far from Fourier-transform limited.  Gaussian transform limited pulses of such duration correspond to a spectral {\blu broadening} of about $15$~GHz~\cite{yariv06}.   The excessive spectral broadening is attributed to laser frequency chirp,  {\blu which requires a proper theoretical description to understand the results of the interference experiment in the presence of time jitter.}

{\blu We focus on the beam splitter and evaluate the coincidence counts registered by the two detectors. The two optical pulses generated by the independent GS laser diodes enter the beam splitter through inputs $a$ and $b$. The electric field at $k=\{a,b\}$ and time $t$ is:}
\begin{equation}
  E_k (t) = \sqrt{I(t-t_k)} \exp\{2 \pi i[\nu (t-t_k) + \beta (t-t_k)^2 + \varphi_k]\},
\label{eq:elfield}
\end{equation}
where $I(t)=\exp\{-t^2/2\tau_p^2\}/(\tau _ p \sqrt{2 \pi})$ {\blu is the temporal profile of the laser pulse, assumed to be Gaussian,} and $\tau_p$ its temporal width; {\blu $\nu$ is the central frequency of the wavepackets; $t_k$ is the temporal distance of wavepacket $k$ from the beam splitter at time $t$;} $\beta$ is a parameter accounting for frequency chirp, {\blu which is about 0.01~ps$^{-2}$ in semiconductor lasers~\cite{agrawal02}; $\varphi_k$ is the (random) electromagnetic phase of the pulses. We also define for later convenience the time delay between the two pulses $\Delta t=t_b - t_a$ and the phase difference $\Delta \varphi=\varphi_b - \varphi_a$.} The time delay can be due to either systematic temporal misalignment or emission time uncertainty.

At the beam splitter, the pulses are aligned to same polarisation and interfere.  The output intensities can be {\blu calculated from Eq.~(\ref{eq:elfield}) and the {\blu beam splitter relations~\cite{loudon00}}. After integrating over the finite response time of the detectors, much longer than the pulse width, and assuming a 50/50 beam splitter}, we obtain for the intensity at the output ports of the beam splitter, $c$ and $d$,
\begin{equation}
I_{c,d} =  1 \pm \cos(\Delta\varphi)
\exp [-\frac {(\Delta t)^2} {8 \tau_p^2}(1+ 16 \beta^2 \tau_p^4)],
\label{eq:Icd}
\end{equation}
\noindent where the +(-) sign is associated with the $c$ ($d$) mode. 
Figure~\ref{fig:fig1}(c) shows example traces of $I_c$ and $I_d$ recorded at the beam splitter output ports using a pair of fast photodiodes {\blu as detectors, after having set} optical attenuation {\blu to 0~dB} and spectral filtering {\blu to pass all frequency components} (see setup in Fig.~\ref{fig:fig2}). Photodiodes record complementary outputs, as a result of energy conservation. Peak intensities fluctuate because of the random phase difference $\Delta \varphi$ in Eq.~(\ref{eq:Icd}).
Nearly complete constructive and destructive interference {\blu  is} {\blu observable because of the} occasional perfect temporal alignment ($\Delta t \approx 0$) {\blu of the two wavepackets}. Since we do not limit the frequency chirp here, the observation of nearly complete interference suggests that {\blu the} two lasers have a similar chirp{\blu ed} profile. {\blu However, as we shall see, this is not sufficient to guarantee a high visibility in two-pulse interference.}

Most often $\Delta t \neq 0$ because of the emission time jitter.  In this case,  frequency chirp will prevent complete interference and hence deteriorate the interference visibility.  The differential phase between two pulses is no longer constant, but evolves as $\Delta \varphi (t) =  \Delta \varphi_0 +  2 \pi \beta  \Delta t \cdot t$.  Crudely speaking, whenever the differential phase evolves by more than $2\pi$, half of the optical wave interferes constructively and the other half destructively, resulting an overall interference visibility approaching zero.

{\blu The average two-pulse interference visibility can be obtained as $V^{(2)}=1-P_{cd}$, where $P_{cd} \propto I_c \cdot I_d$ is the (normalized) coincidence rate seen by the two detectors under the assumption of attenuated intensities~\cite{rarity05}. After averaging over the random phase difference $\Delta\varphi$ and under the experimentally fulfilled condition of pulses attenuated at the single photon level, we obtain the visibility as:}
\begin{equation}
V^{(2)} = \frac{1}{2}\exp [-\frac {(\Delta t)^2} {4 \tau_p^2}(1+ 16 \beta^2 \tau_p^4)].
\label{eq:Pcd}
\end{equation}
{\blu It is worth remarking that the theoretical limit for the interference visibility is 50\% in this case, not 100\%, because attenuated coherent states, not single photons, are interfering at the beam splitter.}

\begin{figure}[tbp]
\centering
\includegraphics[width=.88\columnwidth]{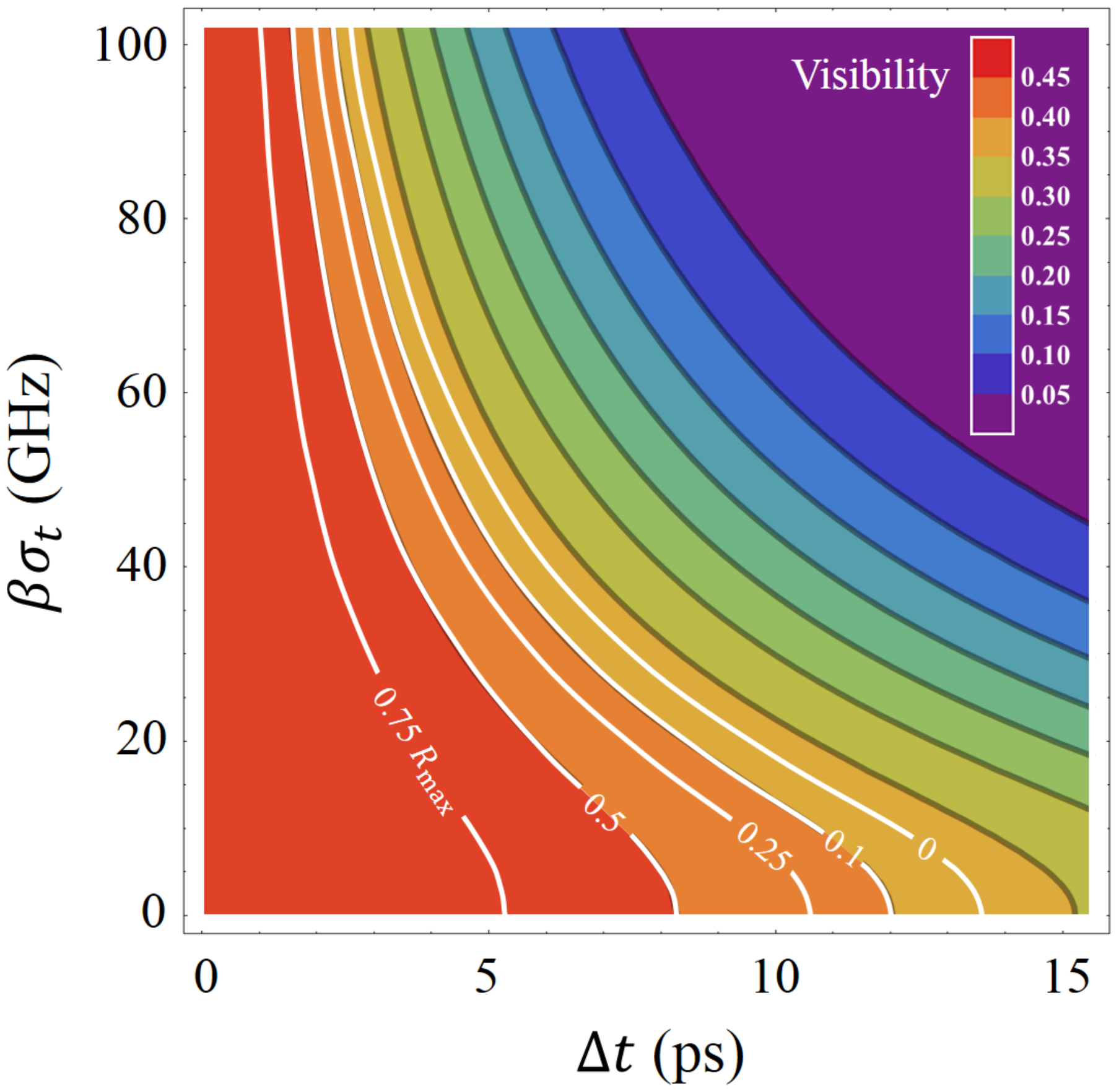}
\caption{Calculated second order interference visibility $V^{(2)}$ as a function of frequency chirp ($\beta \sigma_t$) and temporal misalignment ($\Delta t$) of interfering pulses. We assume a temporal width of 30~ps (FWHM) for laser pulses.  White lines indicate the achievable secure bit rates of MDI-QKD, compared to the rate $R_{max}$ achievable with perfect interference.}
\label{fig:tli}
\newpage
\end{figure}

In Fig.~\ref{fig:tli}, we {\blu plot} the interference visibility as a function of temporal misalignment and frequency chirp. We use $\tau_p = 12.7$~ps, corresponding to an FWHM of 30~ps for a Gaussian wavepacket. When both lasers are perfectly aligned and jitter free ($\Delta t =0$), a visibility of $V^{(2)}=0.5$ is obtained irrespective of the amount of frequency chirp, as expected for perfectly indistinguishable, phase-randomized  weak laser  pulses.
Similarly, in the absence of frequency chirp ($\beta = 0$), the two-pulse interference also exhibits high visibility as long as the temporal misalignment is insignificant as compared with the pulse duration. However, the visibility deteriorates rapidly when both temporal misalignment and frequency chirp are present. With a realistic temporal misalignment ($\Delta t = 10$~ps) and frequency chirp ($\beta \sigma_t = 70$~GHz), the interference visibility drops to $\sim$0.10, a value too low for any practical applications.

As an example, low visibility reduces the secure key rate of MDI-QKD. This is because visibility directly affects the phase error rate in the protocol, thereby increasing
the privacy amplification cost. Using realistic parameters for channel transmission of 0.2~dB/km and measurement efficiency of 30\%,  a maximum secure key rate ($R_{max}$) of the order of 10~kbps can be attained with a GHz-clocked MDI-QKD system over 100 km fiber \cite{error,xu13}. Incidentally, this secure key rate is more than two orders of magnitude greater than what has previously been reported in the literature~\cite{silva13,rubenok13,liu13,tang13mdi}.
However, it will decrease rapidly with deterioration of the visibility.  In Fig.~\ref{fig:tli}, we plot contour lines illustrating the achievable secure key rates at corresponding interference visibilities.
With a slight drop of the visibility from 0.50 to 0.45, the secure key rate is reduced to less than the half of $R_{max}$. It will reduce to around 10\% of $R_{max}$ if the visibility is less than 0.40. When the visibility is lower than 0.37, the generation of a secure key is no longer possible. Hence, high visibility interference is vital to maintain efficient secure key generation in MDI-QKD.

{\blu Having described our theoretical model for two-pulse interference visibility in relation to laser frequency chirp and time jitter, and its effect on the MDI-QKD key rate,} we can now proceed and measure {\blu the real visibility obtained from an MDI-QKD-like setup like the one in Fig.~\ref{fig:fig2}, set in single photon counting mode. Specifically,} both lasers are equally attenuated to $<0.05$ photons per pulse {\blu and superconducting nanowire single-photon detectors with $\sim$5\% quantum efficiency are employed~\cite{error2}}

\begin{figure}[tbp]
\centering
\includegraphics[width=.92\columnwidth]{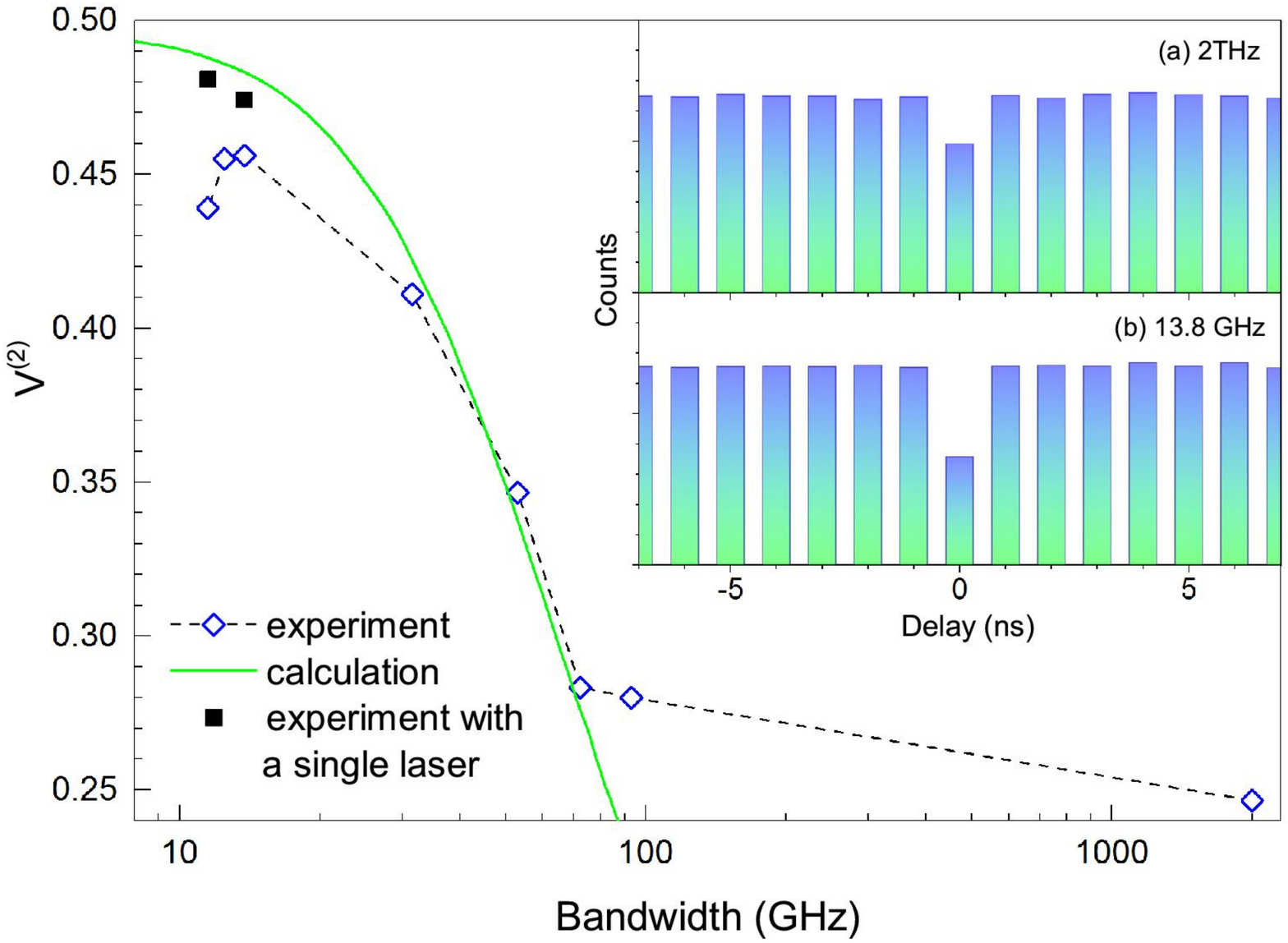}
\caption{The second-order interference visibility $V^{(2)}$ as a function of  filter bandwidth.  Measurements using a single laser and theoretical calculation are also shown.
Insets: Coincidence traces for the Hong-Ou-Mandel interference measurements are shown for two different bandwidths of  (a) 2~THz and (b) 13.8~GHz.
}
\label{fig:HOM}
\newpage
\end{figure}

By setting the filter bandwidth to 2~THz, we record a coincidence histogram as shown in Fig.~\ref{fig:HOM}(a). The suppression at the zero delay corresponds to a visibility of $V^{(2)} = 0.25$.  The visibility is vastly improved to $0.46$ by narrowing the filter bandwidth to 13.8~GHz, as shown in Fig.~\ref{fig:HOM}(b).
We plot the interference visibility as a function of the filter bandwidth in Fig.~\ref{fig:HOM}.  Three different regions can be distinguished in the data.  In the first region with the filter bandwidth greater than 70~GHz,  the visibility improves slowly when the filter narrows.  In this region, the filter rejects only the spontaneous background and side-mode emissions, which are typically three orders of magnitude weaker than the lasing mode. The visibility improves from $V^{(2)}=0.25$  at 2~THz to 0.28 at 72.5~GHz.  Then the visibility improves rapidly when the filter starts to limit the laser bandwidth until reaching a peak visibility of $V^{(2)}=0.46$ at 13.8~GHz. {\blu Note that this bandwidth value is readily obtainable by using appropriately chosen} {\blu off-the-shelf telecom filters~\cite{patel14}}. In the third region,  the visibility starts to deteriorate for filter bandwidths less than 13.8~GHz.

We calculate the visibility as a function of the laser bandwidth using the measured time jitter values and the measured bandwidth-dependent pulse durations. {\blu The resulting theoretical curves fit the experimental data without any free parameter{\blu s}.} As shown in Fig.~\ref{fig:HOM}, the {\blu model} has reproduced the visibility improvement in the intermediate bandwidth region whereas for small bandwidth it shows a considerably higher visibility than actually measured. The discrepancy in the narrow bandwidth region is attributed to imperfection in the measurement setup. We use a single bandpass filter which has a finite back-reflection ratio. {\blu The back-reflected light does not affect the lasers, protected by attenuators and optical isolators, but it can} enter the 50/50 beam splitter {\blu and reach {\blu the} detectors, thus causing accidental coincidences that spoil} the interference visibility. To unveil the truly achievable visibility, we {\blu have} interfered laser pulses emitted by a single laser diode. An asymmetric Mach-Zehnder interferometer is aligned to interfere optical pulses of adjacent clocks~\cite{yuan14}. In this arrangement, the bandpass filter is placed before the interferometer and the back-reflection problem is thus avoided. The results are {\blu also plotted} in Fig.~\ref{fig:HOM}. {\blu This time,} a visibility of $0.48$ is recorded at $11.5$~GHz, {\blu which} {\blu agrees well} with {\blu the predicted} {\blu visibility of} 0.488. The small discrepancy is due to the imperfect splitting ratio in the 50/50 beam-splitter, {\blu which has been measured to be close to 53/47}.
\begin{figure}[t]
\centering
\includegraphics[width=.92\columnwidth]{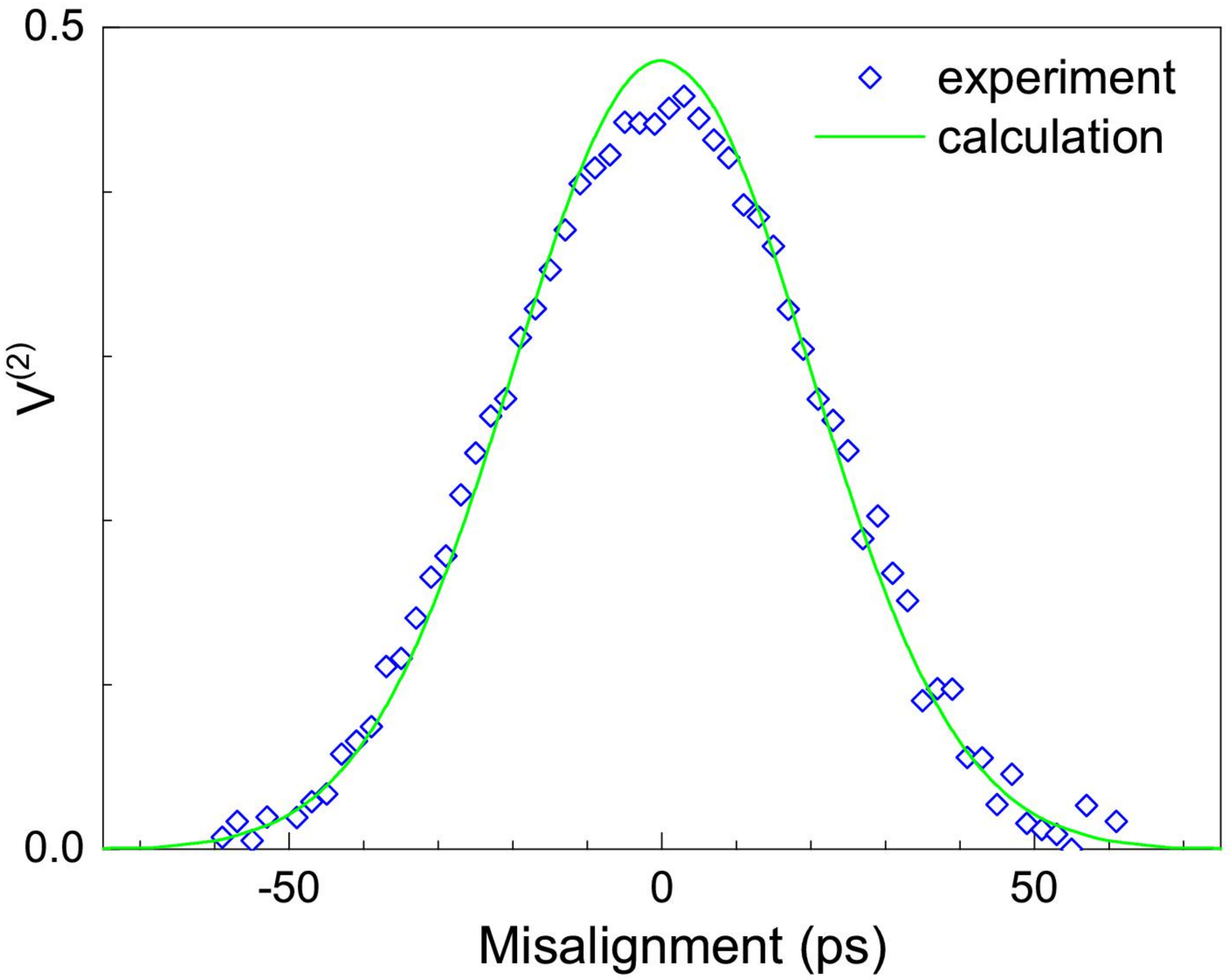}
\caption{Measured (symbols) and simulated (solid line) interference visibility $V^{(2)}$ as a function of temporal misalignment $\Delta t$.}
\label{fig:fig5}
\newpage
\end{figure}

Figure~\ref{fig:fig5} shows the interference visibility as a function of systematic temporal misalignment ($\Delta  t$) between two lasers.  Here, the bandpass filter is fixed to give a bandwidth of 13.8~GHz. On top of laser timing jitter, the systematic misalignment further deteriorates the interference visibility.  In the extreme case of large misalignment  ($|\Delta t|>45$~ps),  the optical pulses have little overlap and the corresponding visibility approaches zero.    Around the optimal delay $\Delta t = 0$, the visibility varies slowly with the temporal misalignment.  At a misalignment of 10~ps, the visibility remains as high as  0.41, a value that is still sufficient for positive key distillation in MDI-QKD.  This temporal tolerance is readily achievable through remote optical synchronisation~\cite{patel12prx}.

Our results {\blu are not limited to MDI-QKD and are} useful to other quantum information applications.
{\blu For instance, a fast random number generator could be envisaged if numbers are assigned to the complementary outcomes shown in Fig.~\ref{fig:fig1}(c), {\blu improving the flexibility of existing solutions based on first-order interference}}~\cite{yuan14,jofre11}.
{\blu Optical interference could also be used} to quantify side-channel information in QKD implementations with multiple light sources. {\blu Currently} information leakage is estimated trough a series of \textit{ad hoc} measurements based on a few known degrees of freedom~\cite{nauerth09}.
{\blu However information leakage from unknown degrees of freedom, sometimes referred to as side channels, cannot be ignored. This security risk could be unveiled by a decrease in the visibility of a multi-source interference experiment.}
{\blu Finally, our demonstration} that Fourier-transform limited pulses are not necessary for {\blu high-visibility interference} may allow weak coherent pulses to be tailored to interfere with quantum light sources, providing a plethora of opportunities, for example, a hybrid quantum relay{\blu ~\cite{lo14}} that bridges weak-pulse QKD and entangled photon pairs{\blu ~\cite{stevenson13}}.

To summarise, we have demonstrated {\blu high-speed} phase-randomised coherent {\blu state} sources {\blu that} can exhibit high visibility in two-pulse interference. {\blu The solution is based on} semiconductor gain-switched laser diodes and {\blu characterized in their temporal and spectral properties.}
{\blu This result is highly beneficial to the recent application of MDI-QKD and to others exploiting similar principles}.
The achieved visibility of 0.46 (Fig.~\ref{fig:HOM}), limited by back-reflection in the filter, {\blu can already guarantee more than 50\% {\blu of} the maximum key rate in MDI-QKD.} Despite {\blu high speed and narrow pulse width}, the interference visibility {\blu obtained} from {\blu frequency-filtered} gain-switched laser diodes is comparable to, or better than, those achieved with optical pulses carved from continuous-wave lasers~\cite{silva13, rubenok13, tang13mdi}.
{\blu This suggests that this cheap and effective solution will play a major role in future quantum-based applications. }


\begin{thebibliography}{99}

\bibitem{gisin02} N. Gisin, G. Ribordy, W. Tittel, and H. Zbinden, \textit{Quantum cryptography}, Rev. Mod. Phys. \textbf{74}, 00145 (2002).

\bibitem{bouwmeester97} D. Bouwmeester, J.-W. Pan, K. Mattle, M. Eibl, H. Weinfurter, and A. Zeilinger, \textit{Experimental quantum teleportation}, Nature \textbf{390}, 575 (1997).

\bibitem{briegel98} H.-J. Briegel, W. D¨ur, J. I. Cirac, and P. Zoller, \textit{Quantum repeaters: The role of imperfect local operations in quantum communication}, Phys. Rev. Lett. \textbf{81}, 5932 (1998).

\bibitem{knill01} E. Knill, R. Laﬂamme, and G. J. Milburn, \textit{A scheme for efficient quantum computation with linear optics}, Nature \textbf{409}, 46 (2001).

\bibitem{hong87} C. Hong, Z. Ou, and L. Mandel, \textit{Measurement of subpicosecond time intervals between two photons by interference}, Phys. Rev. Lett. \textbf{59}, 2044 (1987).

\bibitem{rarity05} J. Rarity, P. Tapster, and R. Loudon, \textit{Non-classical interference between independent sources}, J. Opt. B 7, \textbf{S171} (2005).

\bibitem{kaltenbaek06} R. Kaltenbaek, B. Blauensteiner, M. Zukowski, M. Aspelmeyer, and A. Zeilinger, \textit{Experimental interference of independent photons}, Phy. Rev. Lett. \textbf{96}, 240502 (2006).

\bibitem{beugnon06} J. Beugnon, M. P. Jones, J. Dingjan, B. Darqui´e, G. Messin, A. Browaeys, and P. Grangier, \textit{Quantum interference between two single photons emitted by independently trapped atoms}, Nature \textbf{440}, 779 (2006).

\bibitem{bennett09} A. J. Bennett, R. B. Patel, C. A. Nicoll, D. A. Ritchie,and A. J. Shields, \textit{Interference of dissimilar photon sources}, Nature Phys. \textbf{5}, 715 (2009).

\bibitem{lo05} H. K. Lo, X. F. Ma, and K. Chen, \textit{Decoy state quantum key distribution}, Phys. Rev. Lett. \textbf{94}, 230504 (2005).

\bibitem{wang05} X. B. Wang, \textit{Beating the photon-number-splitting attack in practical quantum cryptography}, Phys. Rev. Lett. \textbf{94}, 230503 (2005).

\bibitem{yuan14} Z. L. Yuan, M. Lucamarini, J. F. Dynes, B. Fr\"{o}hlich, A. Plews, and A. J. Shields, \textit{Robust random number generation using steady-state emission of gain-switched laser diodes}, Appl. Phys. Lett. \textbf{104}, 261112 (2014).

\bibitem{kobayashi14} T. Kobayashi, A. Tomita, and A. Okamoto, \textit{Evaluation of the phase randomness of a light source in quantum-key-distribution systems with an attenuated laser}, Phys. Rev. A \textbf{90}, 032320 (2014).

\bibitem{lutkenhaus09} N. L\"{u}tkenhaus and A. Shields, \textit{Focus on quantum cryptography: theory and practice}, New J. Phys. \textbf{11}, 045005 (2009).

\bibitem{sasaki11} M. Sasaki, M. Fujiwara, H. Ishizuka, W. Klaus, K. Wakui, M. Takeoka, S. Miki, T. Yamashita, Z. Wang, A. Tanaka, K. Yoshino, Y. Nambu, S. Takahashi, A. Tajima, A. Tomita, T. Domeki, T. Hasegawa, Y. Sakai, H. Kobayashi, T. Asai, K. Shimizu, T. Tokura, T. Tsurumaru, M. Matsui, T. Honjo, K. Tamaki, H. Takesue, Y. Tokura, J. F. Dynes, A. R. Dixon, A. W. Sharpe, Z. L. Yuan, A. J. Shields, S. Uchikoga, M. Legr´e, S. Robyr, P. Trinkler, L. Monat, J.-B. Page, G. Ribordy, A. Poppe, A. Allacher, O. Maurhart, T. L¨anger, M. Peev, and A. Zeilinger, \textit{Field test of quantum key distribution in the Tokyo QKD Network}, Opt. Express \textbf{19}, 10387 (2011).

\bibitem{patel12prx} K. A. Patel, J. F. Dynes, I. Choi, A. W. Sharpe, A. R. Dixon, Z. L. Yuan, R. V. Penty, and A. J. Shields, \textit{Coexistence of high-bit-rate quantum key distribution and data on optical fiber}, Phys. Rev. X \textbf{2}, 041010 (2012).

\bibitem{lo12} H.-K. Lo, M. Curty, and B. Qi, \textit{Measurement-device-independent quantum key distribution}, Phys. Rev. Lett. \textbf{108}, 130503 (2012).

\bibitem{silva13} T. F. da Silva, D. Vitoreti, G. Xavier, G. do Amaral, G. Tempor˜ao, and J. von der Weid, \textit{Proof-of-principle demonstration of measurement-device-independent quantum key distribution using polarization qubits}, Phys. Rev. A \textbf{88}, 052303 (2013).

\bibitem{rubenok13} A. Rubenok, J. A. Slater, P. Chan, I. Lucio-Martinez, and W. Tittel, \textit{Real-world two-photon interference and proof-of-principle quantum key distribution immune to detector attacks}, Phys. Rev. Lett. \textbf{111}, 130501 (2013).

\bibitem{liu13} Y. Liu, T.-Y. Chen, L.-J. Wang, H. Liang, G.-L. Shentu, J. Wang, K. Cui, H.-L. Yin, N.-L. Liu, L. Li, X. Ma, J. S. Pelc, M. M. Fejer, C.-Z. Peng, Q. Zhang, and J.-W. Pan, \textit{Experimental measurement-device-independent quantum key distribution}, Phys. Rev. Lett. \textbf{111}, 130502 (2013).

\bibitem{tang13mdi} Z. Tang, Z. Liao, F. Xu, B. Qi, L. Qian, and H.-K. Lo, 	\textit{Experimental demonstration of polarization encoding measurement-device-independent quantum key distribution}, Phys. Rev. Lett. \textbf{112}, 190503 (2014).

\bibitem{dixon08} A. R. Dixon, Z. L. Yuan, J. F. Dynes, A. W. Sharpe, and A. J. Shields, \textit{Gigahertz decoy quantum key distribution with 1 Mbit/s secure key rate}, Opt. Express \textbf{16}, 18790 (2008).

\bibitem{zhao07} Y. Zhao, B. Qi, and H.-K. Lo, \textit{Experimental quantum key distribution with active phase randomization}, Appl. Phys. Lett. \textbf{90}, 044106 (2007).

\bibitem{error3} At least one extra intensity modulator is required to carve the CW light, plus extra random number generator and phase modulator to randomize the phases of the pulses.

\bibitem{agrawal02} G. P. Agrawal,  Fiber-optic communication systems  (John Wiley \& Sons, 2002).

\bibitem{pataca97} D. M. Pataca, P. Gunning, M. L. Rocha, J. K. Lucek, R. Kashyap, K. Smith, D. G. Moodie, R. P. Davey, R. F. Souza, and A. S. Siddiqui, \textit{Gain-switched DFB lasers},  J. Microwav. Optoelectron. \textbf{1}, 46 (1997).

\bibitem{yariv06} A. Yariv and P. Yeh, Photonics: Optical Electronics in Modern Communications (Oxford University Press, 2006).

\bibitem{error4} This does not represent a loss of generality as experimentally it was found that crossing the filter from different directions is tantamount to having different filtering effects on the two beams.

\bibitem{dixon09} A. R. Dixon, J. F. Dynes, Z. L. Yuan, A. W. Sharpe, A. J. Bennett, and A. J. Shields, \textit{Ultrashort dead time of photon-counting InGaAs avalanche photodiodes}, Appl. Phys. Lett. \textbf{94}, 231113 (2009).

\bibitem{loudon00} {\blu R. Loudon, The quantum theory of Light (Oxford University Press, 2000).}

\bibitem{error} We calculate  the secure  bit rate using the decoy  method illustrated in~\cite{xu13} and assuming 4$^\circ$ polarization misalignment in each channel.

\bibitem{xu13} F. Xu, M. Curty, and H. K. Lo, \textit{Practical aspects of measurement-device-independent quantum key distribution}, New  J. Phys. \textbf{15}, 113007 (2013).

\bibitem{error2} Notice that our model holds so long as the detectors are in linear regime, which is the case if the injected average photon number is as small as $0.05$.

\bibitem{patel14} K. A. Patel, J. F. Dynes, M. Lucamarini, I. Choi, A. W. Sharpe, Z. L. Yuan, R. V. Penty, and A. J. Shields, \textit{Quantum key distribution for 10 Gb/s dense wavelength division multiplexing networks}, Appl. Phys. Lett. \textbf{104}, 051123 (2014).

\bibitem{jofre11} M. Jofre,  M. Curty,  F. Steinlechner,  G. Anzolin,  J. P. Torres, M. W. Mitchell,  and V. Pruneri, \textit{True random numbers from amplified quantum vacuum}, Opt. Express \textbf{19}, 20665 (2011).

\bibitem {nauerth09} S. Nauerth, M. F\"{u}rst, T. Schmitt-Manderbach, H. Weier, and H. Weinfurter, \textit{Information leakage via side channels in freespace BB84 quantum cryptography}, New J. Phys. \textbf{11}, 065001 (2009).

\bibitem{lo14} H.-K. Lo, M. Curty, K. Tamaki, \textit{Secure quantum key distribution}, Nat. Photon. \textbf{8}, 595 (2014).

\bibitem{stevenson13} R. M. Stevenson,	J. Nilsson,	A. J. Bennett,	J. Skiba-Szymanska,	I. Farrer,	D. A. Ritchie, and A. J. Shields, \textit{Quantum teleportation using a light-emitting diode}, Nat. Comm. \textbf{4}, 2859 (2013).

\end{thebibliography}
\end{document}